\documentclass[conference]{IEEEtran}
\newtheorem{definition}{Definition}
\newtheorem{theorem}{Theorem}
\usepackage{graphicx}
\newdimen\proofrulebreadth \proofrulebreadth=.05em
\newdimen\proofdotseparation \proofdotseparation=1.25ex
\newdimen\proofrulebaseline \proofrulebaseline=2ex
\newcount\proofdotnumber \proofdotnumber=3
\let\then\relax
\def\hfi{\hskip0pt plus.0001fil}
\mathchardef\squigto="3A3B
\newif\ifinsideprooftree\insideprooftreefalse
\newif\ifonleftofproofrule\onleftofproofrulefalse
\newif\ifproofdots\proofdotsfalse
\newif\ifdoubleproof\doubleprooffalse
\let\wereinproofbit\relax
\newdimen\shortenproofleft
\newdimen\shortenproofright
\newdimen\proofbelowshift
\newbox\proofabove
\newbox\proofbelow
\newbox\proofrulename
\def\shiftproofbelow{\let\next\relax\afterassignment\setshiftproofbelow\dimen0 }
\def\shiftproofbelowneg{\def\next{\multiply\dimen0 by-1 }%
\afterassignment\setshiftproofbelow\dimen0 }
\def\setshiftproofbelow{\next\proofbelowshift=\dimen0 }
\def\setproofrulebreadth{\proofrulebreadth}

\def\prooftree{
\ifnum  \lastpenalty=1
\then   \unpenalty
\else   \onleftofproofrulefalse
\fi
\ifonleftofproofrule
\else   \ifinsideprooftree
        \then   \hskip.5em plus1fil
        \fi
\fi
\bgroup
\setbox\proofbelow=\hbox{}\setbox\proofrulename=\hbox{}%
\let\justifies\proofover\let\leadsto\proofoverdots\let\Justifies\proofoverdbl
\let\using\proofusing\let\[\prooftree
\ifinsideprooftree\let\]\endprooftree\fi
\proofdotsfalse\doubleprooffalse
\let\thickness\setproofrulebreadth
\let\shiftright\shiftproofbelow \let\shift\shiftproofbelow
\let\shiftleft\shiftproofbelowneg
\let\ifwasinsideprooftree\ifinsideprooftree
\insideprooftreetrue
\setbox\proofabove=\hbox\bgroup$\displaystyle 
\let\wereinproofbit\prooftree
\shortenproofleft=0pt \shortenproofright=0pt \proofbelowshift=0pt
\onleftofproofruletrue\penalty1
}

\def\eproofbit{
\ifx    \wereinproofbit\prooftree
\then   \ifcase \lastpenalty
        \then   \shortenproofright=0pt  
        \or     \unpenalty\hfil         
        \or     \unpenalty\unskip       
        \else   \shortenproofright=0pt  
        \fi
\fi
\global\dimen0=\shortenproofleft
\global\dimen1=\shortenproofright
\global\dimen2=\proofrulebreadth
\global\dimen3=\proofbelowshift
\global\dimen4=\proofdotseparation
\global\count255=\proofdotnumber
$\egroup  
\shortenproofleft=\dimen0
\shortenproofright=\dimen1
\proofrulebreadth=\dimen2
\proofbelowshift=\dimen3
\proofdotseparation=\dimen4
\proofdotnumber=\count255
}

\def\proofover{
\eproofbit 
\setbox\proofbelow=\hbox\bgroup 
\let\wereinproofbit\proofover
$\displaystyle
}%
\def\proofoverdbl{
\eproofbit 
\doubleprooftrue
\setbox\proofbelow=\hbox\bgroup 
\let\wereinproofbit\proofoverdbl
$\displaystyle
}%
\def\proofoverdots{
\eproofbit 
\proofdotstrue
\setbox\proofbelow=\hbox\bgroup 
\let\wereinproofbit\proofoverdots
$\displaystyle
}%
\def\proofusing{
\eproofbit 
\setbox\proofrulename=\hbox\bgroup 
\let\wereinproofbit\proofusing
\kern0.3em$
}

\def\endprooftree{
\eproofbit 
  \dimen5 =0pt
\dimen0=\wd\proofabove \advance\dimen0-\shortenproofleft
\advance\dimen0-\shortenproofright
\dimen1=.5\dimen0 \advance\dimen1-.5\wd\proofbelow
\dimen4=\dimen1
\advance\dimen1\proofbelowshift \advance\dimen4-\proofbelowshift
\ifdim  \dimen1<0pt
\then   \advance\shortenproofleft\dimen1
        \advance\dimen0-\dimen1
        \dimen1=0pt
        \ifdim  \shortenproofleft<0pt
        \then   \setbox\proofabove=\hbox{%
                        \kern-\shortenproofleft\unhbox\proofabove}%
                \shortenproofleft=0pt
        \fi
\fi
\ifdim  \dimen4<0pt
\then   \advance\shortenproofright\dimen4
        \advance\dimen0-\dimen4
        \dimen4=0pt
\fi
\ifdim  \shortenproofright<\wd\proofrulename
\then   \shortenproofright=\wd\proofrulename
\fi
\dimen2=\shortenproofleft \advance\dimen2 by\dimen1
\dimen3=\shortenproofright\advance\dimen3 by\dimen4
\ifproofdots
\then
        \dimen6=\shortenproofleft \advance\dimen6 .5\dimen0
        \setbox1=\vbox to\proofdotseparation{\vss\hbox{$\cdot$}\vss}%
        \setbox0=\hbox{%
                \advance\dimen6-.5\wd1
                \kern\dimen6
                $\vcenter to\proofdotnumber\proofdotseparation
                        {\leaders\box1\vfill}$%
                \unhbox\proofrulename}%
\else   \dimen6=\fontdimen22\the\textfont2 
        \dimen7=\dimen6
        \advance\dimen6by.5\proofrulebreadth
        \advance\dimen7by-.5\proofrulebreadth
        \setbox0=\hbox{%
                \kern\shortenproofleft
                \ifdoubleproof
                \then   \hbox to\dimen0{%
                        $\mathsurround0pt\mathord=\mkern-6mu%
                        \cleaders\hbox{$\mkern-2mu=\mkern-2mu$}\hfill
                        \mkern-6mu\mathord=$}%
                \else   \vrule height\dimen6 depth-\dimen7 width\dimen0
                \fi
                \unhbox\proofrulename}%
        \ht0=\dimen6 \dp0=-\dimen7
\fi
\let\doll\relax
\ifwasinsideprooftree
\then   \let\VBOX\vbox
\else   \ifmmode\else$\let\doll=$\fi
        \let\VBOX\vcenter
\fi
\VBOX   {\baselineskip\proofrulebaseline \lineskip.2ex
        \expandafter\lineskiplimit\ifproofdots0ex\else-0.6ex\fi
        \hbox   spread\dimen5   {\hfi\unhbox\proofabove\hfi}%
        \hbox{\box0}%
        \hbox   {\kern\dimen2 \box\proofbelow}}\doll%
\global\dimen2=\dimen2
\global\dimen3=\dimen3
\egroup 
\ifonleftofproofrule
\then   \shortenproofleft=\dimen2
\fi
\shortenproofright=\dimen3
\onleftofproofrulefalse
\ifinsideprooftree
\then   \hskip.5em plus 1fil \penalty2
\fi
}

\ifCLASSINFOpdf
\else
\fi
\begin{document}
%
\title{An Efficient Binary Technique for
Trace Simplifications of Concurrent Programs}


\author{\IEEEauthorblockN{Mohamed A. El-Zawawy\IEEEauthorrefmark{1}
\qquad \qquad\qquad Mohammad N. Alanazi\IEEEauthorrefmark{2}}
\IEEEauthorblockA{\IEEEauthorrefmark{1}\IEEEauthorrefmark{2}College
of Computer and Information Sciences,\\ Al Imam Mohammad Ibn Saud
Islamic University (IMSIU)\\ Riyadh, Kingdom of Saudi Arabia}
\IEEEauthorblockA{\IEEEauthorrefmark{1}Department of Mathematics,
Faculty of Science
\\ Cairo University\\
Giza 12613, Egypt\\
Email\IEEEauthorrefmark{1}: maelzawawy@cu.edu.eg\\
Email\IEEEauthorrefmark{2}: alanazi@ccis.imamu.edu.sa}}


%


\maketitle

\begin{abstract}
Execution of concurrent programs implies frequent switching between
different thread contexts. This property perplexes analyzing and
reasoning about concurrent programs. Trace simplification is a
technique that aims at alleviating this problem via transforming a
concurrent program trace (execution) into a semantically equivalent
one. The resulted trace typically includes less number of context
switches than that in the original trace.

This paper presents a new static approach for trace simplification.
This approach is based on a connectivity analysis that calculates
for each trace-point connectivity and context-switching information.
The paper also presents a novel operational semantics for concurrent
programs. The semantics is used to prove the correctness and
efficiency of the proposed techniques for connectivity analysis and
trace simplification. The results of experiments testing the
proposed technique on problems treated by previous work for trace
simplification are also shown in the paper. The results prove the
efficiency and effectiveness of the proposed method.
\end{abstract}


%
\IEEEpeerreviewmaketitle

\section{Introduction}\label{intro}

Concurrency~\cite{Pacheco11} is becoming a main stream in
programming due to advances in multi-core hardware. Compared to
other programming techniques, debugging and reasoning about
concurrent programs are not easy jobs; in fact they are very
difficult. This is mainly because of the non-deterministic behavior
of their executions. The debugging difficulty was reported by
research~\cite{DimitrovZ11} comparing debugging resources needed for
concurrent and sequential programs where debugging the former was
found to last, on average, (more than twice) longer  than debugging
the latter. The non-deterministic behavior of execution is caused by
non-deterministic thread interleaving at execution time. This makes
reproducing a bug towards analyzing and resolving it, in most cases,
difficult. Much research~\cite{LiaoWCSKLMR13} has been carried out
for smoothing bug reproductions in concurrent programs.

Context switching~\cite{Pacheco11} is a terminology describing
(fine-grained) interleaving
 of different threads. A relatively large number of
context switches in an execution of a concurrent program complicates
its debugging extremely.  This is so as the number of possible
interactions between threads needing to be reasoned about, in order
to understand a trace (an execution), becomes extremely huge by
thread interleaving. Therefore, it is quite helpful to produce an
equivalent execution trace (of a given one) that has less number of
context switches. This results in increasing the interleaving
granularity. One main source of increase in context switches is
thinking sequentially while coding concurrently. Few
attempts~\cite{HuangZ11} were done to produce techniques for
reduction of context switches in executions (traces) of concurrent
programs.

This paper presents a new technique, Binary Trace Reduction
(\textit{BinTrcRed}), for automatic reductions of context switches
in traces (execution instances) of concurrent programs. This
technique  (transformation) produces an equivalent trace to the
given one and hence the produced trace maintains bugs of the
original one. Therefore the resulted simplified trace can be useful
in the debugging process as it removes the burden of reasoning about
unnecessary fine-grained thread interactions. The proposed technique
has the form of a system of inference rules. This has two advantages
over related work. First the system is relatively easy to understand
and to apply as it is simply structured. Secondly, the system
naturally associates each trace simplification process with a
validity proof which has the form a rule derivation in the system.
This proof is required by many applications like proof-carrying
code~\cite{JobredeauxHNF12}.

\textit{BinTrcRed} is based on the result of a connectivity analysis
that is proposed in this paper and that also has the form of a
system of inference rules. The connectivity analysis simply analyzes
a given trace towards complete information about the number of
context switches and trace-joins where switching takes place. Then
based on this information, a sequence of \textit{binary}
replacements between segments (sequence) of statements constituting
the trace are performed by \textit{BinTrcRed} to reduce the number
of context switches. \textit{BinTrcRed} computes a locally optimal
simplification rather than a globally optimal simplification as the
problem was proved to be NP-hard~\cite{JalbertS10}.

Two measures are used to verify the correctness and efficiency of
the proposed technique. The first measure is theoretical and
provides a robust ground for \textit{BinTrcRed}. This is done via
designing an accurate, yet simple, operational semantics for the
model langauge used in this paper. This model is used to state and
prove the correctness and efficiency of \textit{BinTrcRed}. More
specifically, the semantics is used to prove that any resulting
trace by \textit{BinTrcRed} is equivalent (having the same effect on
memory) to the input one and has a number of context switches that
is less than or equal to that in the original trace. The other
measure is experiential results that were carried out to compare the
performance of \textit{BinTrcRed} to a previous technique. Many
parameters were used towards a fair comparison. Experiments confirm
that our technique is faster and more effective than the previous
technique.

\begin{figure}[!t]
\centering \fbox{
\begin{minipage}{8 cm}
  \includegraphics[scale=0.3]{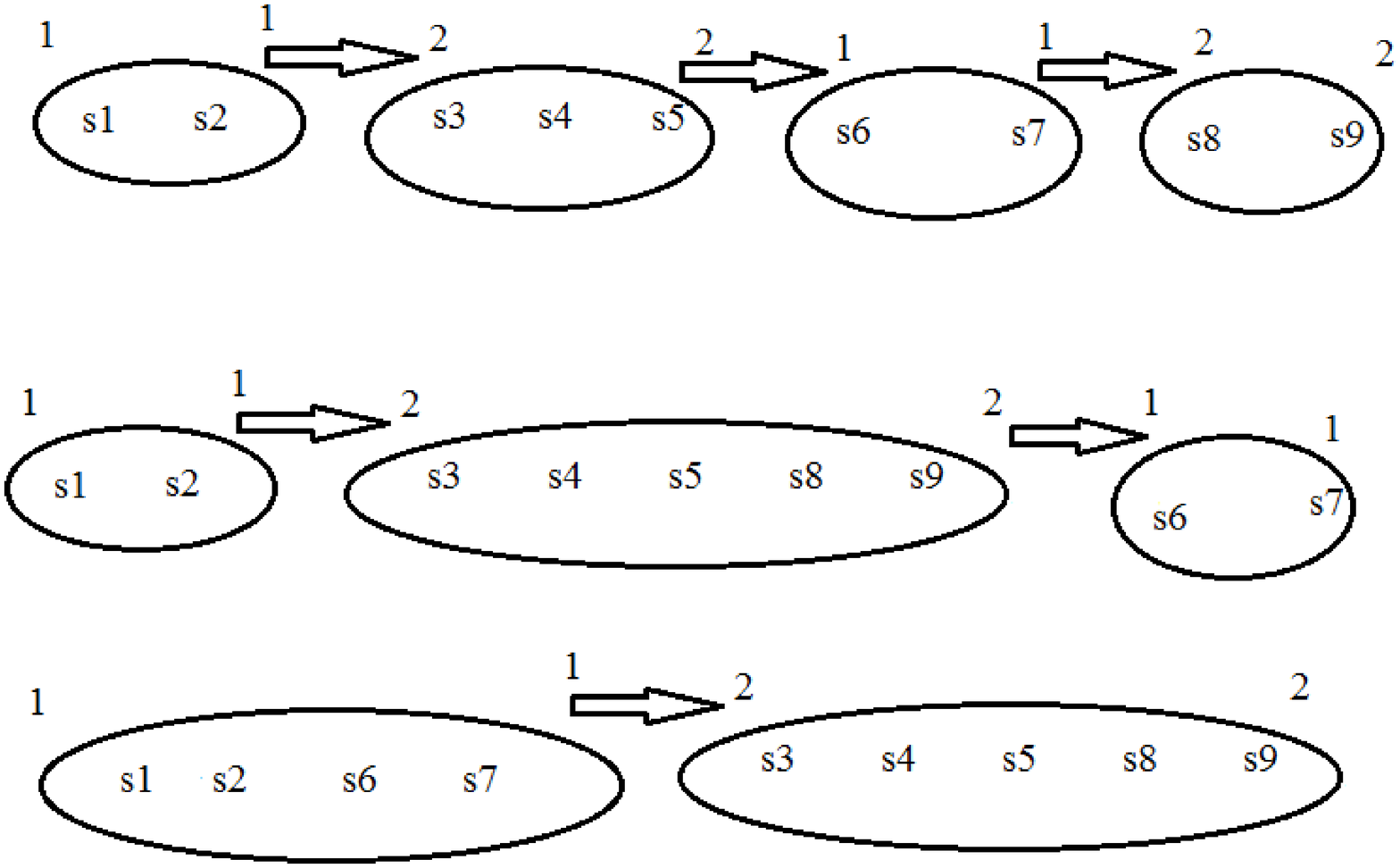}\\
\caption{A Motivating Example.}\label{f0}
\end{minipage}
}
\end{figure}

Figure~\ref{f0} presents a motivating example of the work proposed
in this paper. Assume a concurrent program $P$ that includes $9$
statements distributed between $2$ threads. The upper part of
Figure~\ref{f0} presents a trace of executing this program. This
trace includes $4$ groups of connected statements and $3$ context
switches. For example statements $s3,\ s4$ and $s5$ are connected
and included in thread $2$. After $s5$, a context switch happens to
thread $1$ to execute the connected statements $s6$ and $s7$. First
of all, a robust analysis to accurately collect such connectivity
and switching information is required. Base of the connectivity
information if we replace the third and fourth groups of statements
we get the equivalent trace at the middle of the figure with $2$
context switches. Intuitively, the equivalency is due to the
replacement of unconnected groups of statements. Further
replacements produce the final equivalent trace at the bottom of the
figure with only $1$ context switch. This paper aims at formalizing
a technique that does such replacements. The technique is required
also to associate each such transformation with a correctness proof
that is compact enough for the sake of mobility.

Contributions of this paper are the following:
\begin{enumerate}
\item A new operational approach to
accurately define the semantics of concurrent programs.
\item A connectivity analysis to calculate connectivity and
context switching information in traces of concurrent programs.
\item A new technique to reduce context switches in
traces of concurrent programs.
\end{enumerate}

The outline of this paper is as follows.  Section~\ref{s1} presents
the used model of programming language and the proposed techniques
for connectivity analysis and trace transformation. The semantics
for the langauge constructs together with a formalization for
correctness and efficiency of proposed techniques are shown in
Section~\ref{s2}. Section~\ref{s3} presents the experimental
results. Related and future research are reviewed in
Section~\ref{s4}.

\section{Connectivity Analysis and Trace Transformation}\label{s1}

This section presents our model for a concurrent programming
language. The section also presents two techniques; a connectivity
analysis and a trace transformation reducing number of context
switches in concurrent programs (trace simplification). The section
also uses the language model to introduce a formalization of the
problem of trace simplification (\textit{BinTrcRed}). The language
model includes commands common to languages used to study similar
problems. Figure~\ref{f1} presents the langauge model.

\begin{figure}[t]
\centering \fbox{
\begin{minipage}{8 cm}
{\footnotesize{
\begin{eqnarray*}
& & g\in G=\hbox{Global variable names}\\
& & l\in L=\hbox{Thread local variable names}\\
S \in \hbox{Statements}   &::= &  \hbox{localize}(l,g)\mid
\hbox{Share}(l,g) \mid \hbox{Require}\mid \\ & & \hbox{Release}\mid
\hbox{Duplicate}\mid  \hbox{Initiate}\mid  \hbox{Ready}\mid
\hbox{End}\mid \\&& \hbox{Set1}(g) \mid \hbox{Set0}(g)
\\
T\in \hbox{Thread}   &::= & S\mid T;T^\prime\mid \epsilon.
\\
P\in \hbox{Programs}   &::= & \{T_1\}\dots \{T_n\}.
\end{eqnarray*}
}}\caption{The Language Model.}\label{f1}
\end{minipage}
}
\end{figure}

Some comments on the language model are in order. Two types of
stores are used in the model; global (typical element denoted by
$g$) and local (typical element denoted by $l$). A global store is a
memory location that is accessed by all threads constituting a
concurrent program. A local store is a memory location that is
private for a certain thread. A special global store, $tc$, services
as a counter for the trace. Global stores are meant to facilitate
the communication among a program threads. According to the syntax
of Figure~\ref{f1}, each thread consists of a sequence of
statements. Due to the use of a global trace counter, standing
alone, each thread is deterministic.

Towards a rich, yet simple, langauge model the syntax of our
language includes the following statements:
\begin{itemize}
    \item \textit{Share(l,g)}: copying the value of $g$ into $l$. (Java's read command)
    \item \textit{localize(l,g)}:  copying the value of $l$ to $g$.
    (Java's write command)
    \item \textit{Require}: meaning that its hosting thread requires a
    lock. (Java's lock command)
    \item \textit{Release}: meaning that its hosting thread releases a
    lock. (Java's unlock command)
    \item \textit{Duplicate}: meaning that its hosting thread
    duplicates itself. (Java's fork command)
    \item \textit{Initiate}: meaning that the execution of its hosting
    thread is initiated immediately after the completion of another
    thread. (Java's join command)
    \item \textit{Ready}: meaning that its hosting thread is ready
    for execution. (Java's start command)
    \item \textit{End}: marking the end of a thread. (Java's exit command)
    \item \textit{Set1(g)}: setting the value of $g$ to $1$. (Java's signal command)
    \item \textit{Set0(g)}: waiting $g$ to become $1$ to set it to $0$
    again. (Java's wait command)
\end{itemize}

\begin{definition}\label{trace}
Let $P=\{T_1\}\dots \{T_n\}$ be a program and  suppose that $T_i$
has $n_i$ statements (i.e. $T_i={S_1^i;\dots; S_{n_i}^i}$). Then
\begin{enumerate}
    \item $N_P=\sum_i n_i$.

    \item $S_P=\{S_j^i\mid 1\le i\le n\ \&\ 1\le j\le
n_i\}$

    \item A faithful map $\delta_P$ for the program $P$ is a one-to-one
map\[\delta_P:\{1,\dots,N_P\}\rightarrow S_P\] satisfying the
following condition:

$u,v\in \{1,\dots,N_P\},\ \delta_P(u)=S^i_{q_1},\hbox{ and }
\delta_P(v)=S^i_{q_2}\Longrightarrow q_1< q_2.$

\item The trace, $t_{\delta_P}$, of
$\delta_P$ is the sequence
$\delta_P(1);\delta_P(2);\ldots;\delta_P(N_P)$.

\item For $u\in \{1,\dots,N_P\}$, suppose  $\delta_P(u)=S^i_{q_1}$.
Then
\[th_{\delta_P}:\{1,\dots,N_P\}\rightarrow\{1,\dots,n\};s\mapsto i.\]

\item For $u,v\in \{1,\dots,N_P\}, $
\[\hbox{diff}(u,v)=\left\{
\begin{array}{ll}
0, & th_{\delta_P}(u)=th_{\delta_P}(v)\hbox{;} \\
1, & \hbox{otherwise.}
\end{array}
\right.
\]

\item $CS(t_{\delta_P})=\sum_{s=1}^{N_P-1} \hbox{diff}(s,s+1) $.
\end{enumerate}
\end{definition}

Definition~\ref{trace} introduces concepts necessary to introduce
results of the paper and to formalize the problem of trace
simplification. Some comments on the definition above are in order.
The number and the set of all statements in all threads of a
concurrent program $P$ are denoted by $N_P$ and $S_P$, respectively.
We assume that each program statement is superscribed with its
thread number. Hence when necessary, a trace is denoted by
$S_1^{i_1},\ldots,S_{N_P}^{i_{N_P}},$ where
$1\le,i_1,\ldots,i_{N_P}\le n$. A faithful map (denoted by
$\delta_P$) of a concurrent program ,$P$, is a map that orders the
program statements in way that respects the inner order of each
thread. Hence each trace (denoted by $t_{\delta_P}$) can be realized
as the image of a faithful map $\delta_P$. For a faithful map
$\delta_P(u)$, the map $th_{\delta_P}$ calculates for a given
position in the trace, the ID of the thread that hosts the statement
occupying the position. Using the map $th_{\delta_P}$ of a faithful
map $\delta_P(u)$, the map $\hbox{diff}(u,v)$ decides wether
locations number $u$ and $v$ of the trace are hosting statements of
the same thread. Therefore the summation $\sum_{s=1}^{N_P-1}
\hbox{diff}(s,s+1)$ is the number of the context switches of the
trace in hand.

\begin{definition}\label{connect}
For a trace $S_1^{i_1},\ldots,S_{N_P}^{i_{N_P}}$ of a program
$P=\{T_1\}\dots \{T_n\}$,
\begin{itemize}
    \item $C^P_1=\{(S_1,S_2)\mid \exists 1\le i\le n,
    1\le j \le n_i.\ S_1=S_j^i\hbox{ and } S_2=S^i_{j+1}\}.$
    \item $C^P_2=\{(\hbox{Release}^i,\hbox{Require}^i),
    (\hbox{Duplicate}^i,\hbox{Read}^i),\\(\hbox{End}^i,\hbox{Initiate}^i),
    (\hbox{Set1}^i(g),\hbox{Set0}^{i^\prime}(g))\mid 1\le i,i^\prime\le
    n\}.$
    \item $C^P_3=\{(\hbox{localize}^i(l,g),\hbox{Share}^j(l^\prime,g)),
    (\hbox{Share}^i(l,g),\\\hbox{localize}^j(l^\prime,g)),
    (\hbox{Share}^i(l,g),\hbox{Share}^j(l^\prime,g))\mid i\not =
    j\}.$
    \item The connectivity set: \[C^P=C_1^P\cup C_2^P\cup C_3^P.\]
    \item The map, connect, measuring connectivity of statements in a trace is
    defined as following:
    \[\hbox{connect}(S_1,S_2)=\left\{
                         \begin{array}{ll}
                           1, & (S_1,S_2)\in C^P; \\
                           0, & \hbox{otherwise.}
                         \end{array}
                       \right.
    \]
\end{itemize}
\end{definition}

\begin{definition}\label{atrace}
For a trace $S_1,\ldots,S_{N_P}$ of a program $P=\{T_1\}\dots
\{T_n\}$, an annotated trace is a sequence:

$(s_1^0,s_2^0,t_1^0,t_2^0)\ S_1\ (s_1^1,s_2^1,t_1^1,t_2^1)\ S_2\
(s_1^2,s_2^2,t_1^2,t_2^2) \ldots S_{N_P}\\
(s_1^{N_P},s_2^{N_P},t_1^{N_P},t_2^{N_P}),$

such that $\forall u\in \{1,\dots,N_P\},\ s_1^u,s_2^u\in
\{1,\dots,N_P\}
    \hbox{ and } t_1^u,t_2^u\in \{1,\dots,n\}.$
    \end{definition}

The conditions under which two statements of a concurrent program
are considered connected are presented in~Definition~\ref{connect}.
There are three types of pairs of connected statements in a trace of
a program $P$. The three types are the following. Paris of
contiguous statements of the same thread are grouped in the set
$C_1^P$. The set $C_2^P$ collects pairs of contiguous concurrent
statements of various threads accessing the same global variable.
Pairs of contiguous conflicting concurrent statements are grouped in
the set $C_3^P$. The set of all pairs of connected statements is
denoted by $C^P$. The map $connect(S_1,S_2)$ is binary-valued and
decides connectivity of $S_1$ and $S_2$ using the set $C^P$. To
illustrate Definition~\ref{atrace}, it is necessary to recall that
each trace consists of contiguous segments of statements such that
inside each segment contiguous statements are connected. In an
extreme case, each segment includes only one statement. For a trace,
Definition~\ref{atrace} introduces the concept of an annotated trace
which is a trace whose join-points are annotated with connectivity
information. For a join-point $i$, this information is a quadrable
$(s_1^i,s_2^i,t_1^i,t_2^i)$ where:
\begin{itemize}
    \item the number of the first member in the segment including
    the statement $S_i$ is denoted by $s_1^i$,
    \item the number of the last member in the segment including
    the statement $S_i$ is denoted by $s_2^i$,
    \item the thread ID of the first member in the segment including
    the statement $S_i$ is denoted by $t_1^i$, and
    \item the thread ID of the last member in the segment including
    the statement $S_i$ is denoted by $t_2^i$.
\end{itemize}

\begin{figure}[t]
\centering \fbox{
\begin{minipage}{8cm}
{\footnotesize{
\[
\begin{prooftree}
 \justifies
S_u: (0,0,0,0)\rightarrow (u,u,th_{\delta_P}(u),th_{\delta_P}(u))
\thickness=0.08em\using{(\hbox{tr}_1)}
\end{prooftree}
\]
\[
\begin{prooftree}
\hbox{connect}(\delta_P(u-1),\delta_P(u))=0
 \justifies
S_u: (s_1,s_2,t_1,t_2)\rightarrow
(u,u,\hbox{th}_{\delta_P}(u),\hbox{th}_{\delta_P}(u))
\thickness=0.08em\using{(\hbox{tr}_2)}
\end{prooftree}
\]
\[
\begin{prooftree}
\hbox{connect}(\delta_P(u-1),\delta_P(u))=1
 \justifies
S_u: (s_1,s_2,t_1,t_2)\rightarrow
(s_1,u,t_1,\hbox{th}_{\delta_P}(u))
\thickness=0.08em\using{(\hbox{tr}_3)}
\end{prooftree}
\]
\[
\begin{prooftree}
\begin{tabular}{l}
$S_1: (s_1,s_2,t_1,t_2)\rightarrow
(s_1^{\prime\prime},s_2^{\prime\prime},t_1^{\prime\prime},t_2^{\prime\prime})$\\
$S_2, \ldots,\ S_q:
(s_1^{\prime\prime},s_2^{\prime\prime},t_1^{\prime\prime},t_2^{\prime\prime})
\rightarrow (s_1^\prime,s_2^\prime,t_1^\prime,t_2^\prime)$
\end{tabular}
  \justifies
S_1,\ S_2, \ldots,\ S_q: (s_1,s_2,t_1,t_2)\rightarrow
(s_1^\prime,s_2^\prime,t_1^\prime,t_2^\prime)
\thickness=0.08em\using{(\hbox{tr}_4)}
\end{prooftree}
\]
}} \caption{Rules for Connectivity Analysis.}\label{f2}
\end{minipage}
}
\end{figure}

Figure~\ref{f2} presents the connectivity analysis in the form of a
system of inference rules. For a given trace $S_1,\ldots,S_{N_P}$ of
a program $P=\{T_1\}\dots \{T_n\}$, the idea is to use the rules to
find a quadrable $(s_1,s_2,t_1,t_2)$ such that \[S_1,\ S_2, \ldots,\
S_{N_P}: (0,0,0,0)\rightarrow (s_1,s_2,t_1,t_2)\] is derivable in
the system. If such derivation exists, then an annotated trace (in
the sense of Definition~\ref{atrace} above) can be easily built from
the derivation. The obtained annotated trace includes all the
necessary connectivity information for embarking on reducing the
number of context switches. The precondition of $(\hbox{tr}_3),\
\hbox{connect}(\delta_P(u-1),\delta_P(u))=1$ requires that the
current statement is connected to its prior one. In this case the
current statement is attached to the segment of its prior statement
by letting the information in  the next join-point to be
$(s_1,u,t_1,\hbox{th}_{\delta_P}(u))$.

It is quite important to note that although the proposed
connectivity analysis seems to cost $O(2^n)$, this is not the case
as the method does not actually ensure the connectivity for all
pairs of statements. However, the proposed method ensures
connectivity of (roughly) all join points of the input program.

\begin{figure}[t]
\centering \fbox{
\begin{minipage}{8.6cm}
{\footnotesize{
\[
\begin{prooftree}
\justifies
\begin{tabular}{l}
$(s_1^{u-1},s_2^{u-1},t_1^{u-1},t_2^{u-1})\ S_u\
 (s_1^u,s_2^u,t_1^u,t_2^u)$  \\
$\Longrightarrow\ (s_1^{u-1},s_2^{u-1},t_1^{u-1},t_2^{u-1})\ S_u\
 (s_1^u,s_2^u,t_1^u,t_2^u)$
\end{tabular}
\thickness=0.08em\using{(\hbox{base}_0)}
\end{prooftree}
\]
\[
\begin{prooftree}
t_2^{u-1}\not=t_1^{u+1}\justifies
\begin{tabular}{l}
$(s_1^{u-1},s_2^{u-1},t_1^{u-1},t_2^{u-1})\ S_u\
 (s_1^u,s_2^u,t_1^u,t_2^u)\ S_{u+1}$  \\
$ (s_1^{u+1},s_2^{u+1},t_1^{u+1},t_2^{u+1})\Longrightarrow\
(s_1^{u-1},s_2^{u-1},t_1^{u-1},t_2^{u-1})\ S_u$  \\
$(s_1^u,s_2^u,t_1^u,t_2^u)\ S_{u+1}\
 (s_1^{u+1},s_2^{u+1},t_1^{u+1},t_2^{u+1})$
\end{tabular}
\thickness=0.08em\using{(\hbox{base}_1)}
\end{prooftree}
\]
\[
\begin{prooftree}
t_2^{u-1}=t_1^{u+1}\justifies
\begin{tabular}{l}
$(s_1^{u-1},s_2^{u-1},t_1^{u-1},t_2^{u-1})\ S_u\
 (s_1^u,s_2^u,t_1^u,t_2^u)\ S_{u+1}$  \\
$ (s_1^{u+1},s_2^{u+1},t_1^{u+1},t_2^{u+1})\Longrightarrow\
(s_1^{u-1},s_2^{u-1},t_1^{u-1},t_2^{u-1})$  \\
$S_{u+1}\ (\underline{\textbf{s}_1^{u-1}},s_2^{u+1},
 \underline{\textbf{t}_1^{u-1}},t_2^{u+1})\ S_u\
 (s_1^u,s_2^u,t_1^u,t_2^u)$
\end{tabular}
\thickness=0.08em\using{(\hbox{base}_2)}
\end{prooftree}
\]
\[
\begin{prooftree}
\begin{tabular}{l}
$N_q>2\qquad v=\lceil \mu/2\rceil$   \\
 $(s_1^0,s_2^0,t_1^0,t_2^0)\ S_1\
 (s_1^1,s_2^1,t_1^1,t_2^1)\ S_2\ \ldots\ S_v\
 (s_1^v,s_2^v,t_1^v,t_2^v)$  \\
 $\Longrightarrow\ (s_1^{0\prime},s_2^{0\prime},
 t_1^{0\prime},t_2^{0\prime})\ S_1^\prime\
 (s_1^{1\prime},s_2^{1\prime},t_1^{1\prime},t_2^{1\prime})\ S_2^\prime\
 \ldots $  \\ $ S_v^\prime\
 (s_1^{v\prime},s_2^{v\prime},t_1^{v\prime},t_2^{v\prime})$ \\
  $(s_1^v,s_2^v,t_1^v,t_2^v)\ S_{v+1}\
 (s_1^{v+1},s_2^{v+1},t_1^{v+1},t_2^{v+1})\ S_{v+2}\ \ldots$  \\
$ S_{\mu}
 (s_1^{\mu},s_2^{\mu},t_1^{\mu},t_2^{\mu})\Longrightarrow\
 (s_1^{v\prime},s_2^{v\prime},t_1^{v\prime},
  t_2^{v\prime})\ S_{v+1}^\prime$ \\
  $ (s_1^{v+1\prime},s_2^{v+1\prime},t_1^{v+1\prime},t_2^{v+1\prime})
 \ S_{v+2}^\prime\ \ldots\
 \ S_{\mu}^\prime\
 (s_1^{\mu\prime},s_2^{\mu\prime},t_1^{\mu\prime},t_2^{\mu\prime})$
\end{tabular}
\justifies
\begin{tabular}{l}
$(s_1^0,s_2^0,t_1^0,t_2^0)\ S_1\
 (s_1^1,s_2^1,t_1^1,t_2^1)\ S_2\ \ldots\ S_{\mu}\
 (s_1^{\mu},s_2^{\mu},t_1^{\mu},t_2^{\mu})$   \\
$\Longrightarrow\left\{
\begin{array}{ll}
(s_1^{0\prime},s_2^{0\prime},t_1^{0\prime},t_2^{0\prime})\
S_{v+1}^\prime\ \ldots S_{\mu}^\prime\
(\underline{\textbf{s}_1^{0\prime}},s_2^{\mu\prime},
, &  \\
\underline{\textbf{t}_1^{0\prime}},
 t_2^{\mu\prime})\ S_1^\prime\ \ldots\ S_v^\prime\ (s_1^{v\prime},s_2^{v\prime},
t_1^{v\prime},t_2^{v\prime}) & \hbox{if } t_2^0=t_1^{\mu\prime}; \\
 (s_1^0,s_2^0,t_1^0,t_2^0)\ S_1\
 (s_1^1,s_2^1,t_1^1,t_2^1)\ S_2\ \ldots  & \\
 S_{\mu}\
 \ (s_1^{\mu},s_2^{\mu},t_1^{\mu},t_2^{\mu}), \hbox{otherwsie.}&
                  \end{array}
                \right.$
\end{tabular}
\thickness=0.08em\using{(\hbox{S})}
\end{prooftree}
\]
}} \caption{\textit{BinTrcRed}: Rules for Context Switching
Reduction.}\label{f3}
\end{minipage}
}
\end{figure}

Figure~\ref{f3} presents \textit{BinTrcRed}, the main technique of
the paper for reducing number of context switches in concurrent
program. The technique has the form of a system of inference rules.
The technique builds on the results of the connectivity analysis
introduced above. The rule $(\hbox{base}_0)$ expresses the fact that
the transformation of a single statement is the statement itself
again. For the annotated trace

$(s_1^{u-1},s_2^{u-1},t_1^{u-1},t_2^{u-1})\ S_u\
 (s_1^u,s_2^u,t_1^u,t_2^u)\ S_{u+1}\\
 (s_1^{u+1},s_2^{u+1},t_1^{u+1},t_2^{u+1}),$

if $t_2^{u-1}\not=t_1^{u+1}$, then
 switching the two statements $S_u$ and $S_{u+1}$ would not
 reduce the number of context switches. Therefore as formalized in
 the rule $(\hbox{base}_1)$, the transformation of the trace above
 is the same trace again. However if $t_2^{u-1}=t_1^{u+1}$, then
 switching the two statements $S_u$ and $S_{u+1}$ would
 reduce the number of context switches in the trace by one. This is formalized in
 the rule $(\hbox{base}_2)$. For a longer annotated trace, the rule
 $(\hbox{S})$ breaks the trace into two sub-traces and applies the system
 on each sub-trace. Then the rule switches the two obtained sub-traces
 only if their switching would reduce the number of context switches by
 $1$.

Theorem~\ref{lessCS} states that the number of context switching in
a trace resulted from the transformation system above,
\textit{BinTrcRed}, is less than or equal that number in the ordinal
trace. A straightforward structure induction on rules of
Figure~\ref{f3} proves the theorem.

\begin{theorem}\label{lessCS}
Let $P=\{T_1\}\dots \{T_n\}$ be a program and  suppose that $T_i$
has $n_i$ statements (i.e. $T_i={S_1^i;\dots; S_{n_i}^i}$). Suppose
that $\delta_P$ is a trace for $P$ with annotation:

$(s_1^0,s_2^0,t_1^0,t_2^0)\delta_P(1)(s_1^1,s_2^1,t_1^1,t_2^1)
\delta_P(2)(s_1^2,s_2^2,t_1^2,t_2^2) \ldots \\
\delta_P(N_P)(s_1^{N_P},s_2^{N_P},t_1^{N_P},t_2^{N_P}),$

obtained using the analysis technique of Figure~\ref{f2}. Suppose
that this trace is transomed using \textit{BinTrcRed}
(Figure~\ref{f3}):

$(s_1^0,s_2^0,t_1^0,t_2^0)\delta_P(1)(s_1^1,s_2^1,t_1^1,t_2^1)
\ldots \delta_P(N_P)\\(s_1^{N_P},s_2^{N_P},t_1^{N_P},t_2^{N_P})
\Longrightarrow
(s_1^{0\prime},s_2^{0\prime},t_1^{0\prime},t_2^{0\prime})\delta^\prime_P(1)\\
(s_1^{1\prime},s_2^{1\prime},t_1^{1\prime},t_2^{1\prime}) \ldots
\delta^\prime_P(N_P)
(s_1^{N_P\prime},s_2^{N_P^\prime},t_1^{N_P^\prime},t_2^{N_P^\prime}).$

Then $CS(t_{\delta_P^\prime})\le CS(t_{\delta_P}).$
\end{theorem}

\section{Semantics Based Correctness Formalization}\label{s2}
This section presents a novel semantics for trace executions in
concurrent programming languages. The proposed semantics is
operational and consists of a set of states and a transition
relation between the states. A state is a triple $(\gamma,L,W)$,
where $\gamma$ captures the contents of local and global variables,
$L$ is the set of threads requiring looks at that point of
execution, and $W$ is the set of global variables being watched by
the command $\hbox{Set0}$. Definition~\ref{state} formalizes the
state definition.

\begin{definition}\label{state}
\begin{itemize}
    \item Local locations
    of thread $i$ are $L_i=\{l^i_1,l^i_2,\ldots\}$.
    \item A special global variable is the trace counter denoted by
    $tc$.
    \item A variable state $\gamma$ is a partial map from
    $G\cup \cup_i L_i$ to the
    set of integers.
    \item A trace state is a triple $(\gamma,L,W)$; $L$
    denotes a list of threads requiring a lock and $W$ denotes the
    set of global variables being watched by the  statement "\textit{Set0}".
\end{itemize}
\end{definition}

\begin{figure}[t]
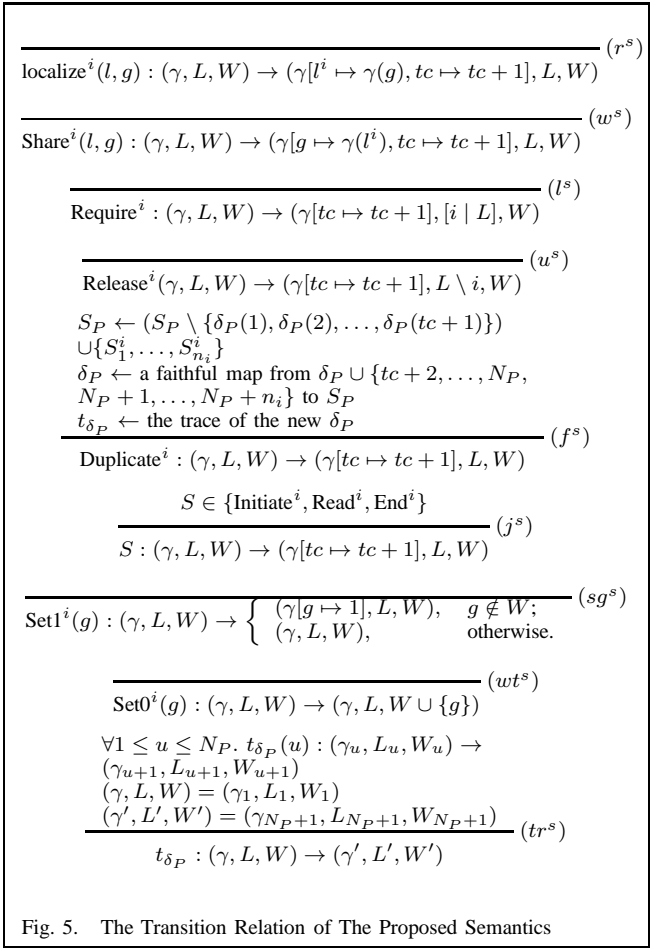

\centering \fbox{
\begin{minipage}{8.1cm}
{\footnotesize{
\[
\begin{prooftree}
 \justifies
\hbox{localize}^i(l,g):(\gamma,L,W)\rightarrow (\gamma[l^i\mapsto
\gamma(g),tc\mapsto tc+1],L,W) \thickness=0.08em\using{(r^s)}
\end{prooftree}
\]\[
\begin{prooftree}
 \justifies
\hbox{Share}^i(l,g):(\gamma,L,W)\rightarrow (\gamma[g\mapsto
\gamma(l^i),tc\mapsto tc+1],L,W) \thickness=0.08em\using{(w^s)}
\end{prooftree}
\]
\[
\begin{prooftree}
 \justifies
\hbox{Require}^i:(\gamma,L,W)\rightarrow (\gamma[tc\mapsto
tc+1],[i\mid L],W) \thickness=0.08em\using{(l^s)}
\end{prooftree}
\]
\[
\begin{prooftree}
 \justifies
\hbox{Release}^i(\gamma,L,W)\rightarrow (\gamma[tc\mapsto
tc+1],L\setminus i,W) \thickness=0.08em\using{(u^s)}
\end{prooftree}
\]
\[
\begin{prooftree}
\begin{tabular}{l}
$S_P\leftarrow (S_P\setminus \{\delta_P(1),\delta_P(2),\ldots,
\delta_P(tc+1)\})$\\ $\cup \{S^i_1,\dots,S^i_{n_i}\}$ \\
$\delta_P\leftarrow$ a faithful map from $
\delta_P\cup\{tc+2,\dots,N_P,$\\ $N_P+1,\ldots,N_P+n_i\} $ to $S_P$\\
$t_{\delta_P}\leftarrow $ the trace of the new $\delta_P$
\end{tabular}
\justifies \hbox{Duplicate}^i:(\gamma,L,W)\rightarrow
(\gamma[tc\mapsto tc+1],L,W) \thickness=0.08em\using{(f^s)}
\end{prooftree}
\]
\[
\begin{prooftree}
 S\in\{\hbox{Initiate}^i,\hbox{Read}^i,\hbox{End}^i\} \justifies
S:(\gamma,L,W)\rightarrow (\gamma[tc\mapsto tc+1],L,W)
\thickness=0.08em\using{(j^s)}
\end{prooftree}
\]
\[
\begin{prooftree}
\justifies \hbox{Set1}^i(g):(\gamma,L,W)\rightarrow\left\{
 \begin{array}{ll}
(\gamma[g\mapsto 1],L,W) , & g\notin W \hbox{;} \\
(\gamma,L,W) , & \hbox{otherwise.}
 \end{array}
 \right.
 \thickness=0.08em\using{(sg^s)}
\end{prooftree}
\]
\[
\begin{prooftree}
\justifies \hbox{Set0}^i(g):(\gamma,L,W)\rightarrow
(\gamma,L,W\cup\{g\}) \thickness=0.08em\using{(wt^s)}
\end{prooftree}
\]
\[
\begin{prooftree}
\begin{tabular}{l}
$\forall 1\le u\le N_P.\
t_{\delta_P}(u):(\gamma_u,L_u,W_u)\rightarrow $\\ $
(\gamma_{u+1},L_{u+1},W_{u+1})$  \\
$(\gamma,L,W)=(\gamma_1,L_1,W_1)$   \\
$(\gamma^\prime,L^\prime,W^\prime)=(\gamma_{N_P+1},L_{N_P+1},W_{N_P+1})$
\end{tabular} \justifies t_{\delta_P}:(\gamma,L,W)\rightarrow
(\gamma^\prime,L^\prime,W^\prime) \thickness=0.08em\using{(tr^s)}
\end{prooftree}
\]
}} \caption{The Transition Relation of The Proposed
Semantics}\label{f4}
\end{minipage}
}
\end{figure}

The transition relation of the proposed operational semantics, in
the form of a system of inference rules, is shown in
Figure~\ref{f4}. Some comments are in order. The rule $(f^s)$
simulates the semantics of the statement $\hbox{Duplicate}^i$. This
is done via adding statements of thread $i$ into the set $S_P$ of
all statements of the program $P$ after removing the already
executed statements from $S_P$. The remaining trace is replaced with
the new trace corresponding to a faithful map for the new set of all
statements.

Theorem~\ref{equivalentS} formalizes the correctness of the
transformation technique, \textit{BinTrcRed}, proposed in the
previous section. This is done using the operational semantics
detailed above. The proof of the theorem is built using structure
induction on transformation and semantics rules.

\begin{theorem}\label{equivalentS}
Let $P=\{T_1\}\dots \{T_n\}$ be a program and  suppose that $T_i$
has $n_i$ statements (i.e. $T_i={S_1^i;\dots; S_{n_i}^i}$). Suppose
that $\delta_P$ is a trace for $P$ with
annotation:

$(s_1^0,s_2^0,t_1^0,t_2^0)\delta_P(1)(s_1^1,s_2^1,t_1^1,t_2^1)
\delta_P(2)s_1^2,s_2^2,t_1^2,t_2^2) \ldots \\
\delta_P(N_P)(s_1^{N_P},s_2^{N_P},t_1^{N_P},t_2^{N_P}),$

obtained using the analysis technique of Figure~\ref{f2}. Suppose
that this trace is transomed using \textit{BinTrcRed}:

$(s_1^0,s_2^0,t_1^0,t_2^0)\delta_P(1)(s_1^1,s_2^1,t_1^1,t_2^1)
\ldots \delta_P(N_P)\\(s_1^{N_P},s_2^{N_P},t_1^{N_P},t_2^{N_P})
\Longrightarrow
(s_1^{0\prime},s_2^{0\prime},t_1^{0\prime},t_2^{0\prime})\delta^\prime_P(1)\\
(s_1^{1\prime},s_2^{1\prime},t_1^{1\prime},t_2^{1\prime}) \ldots
\delta^\prime_P(N_P)
(s_1^{N_P\prime},s_2^{N_P^\prime},t_1^{N_P^\prime},t_2^{N_P^\prime}).$

Suppose that for some $(\gamma,L,W)$, \[
t_{\delta_P}:(\gamma,L,W)\rightarrow
(\gamma^\prime,L^\prime,W^\prime)\] and \[
t_{\delta_P^\prime}:(\gamma,L,W)\rightarrow
(\gamma^{\prime\prime},L^{\prime\prime},W^{\prime\prime}).\]
Then\[(\gamma^\prime,L^\prime,W^\prime)=
(\gamma^{\prime\prime},L^{\prime\prime},W^{\prime\prime}).\]
\end{theorem}

\section{Implementation and Evaluation}\label{s3}

In order to investigate the effectiveness and efficiency of
\textit{BinTrcRed}, several experiments were performed on an
implementation of our proposed technique. \textit{BinTrcRed} was
implemented as a prototype tool for multithreaded Java programs. The
tool includes six phases. The first phase calculates the number of
context switches in the given trace of execution. The second phase
applies the connectivity analysis (Figure~\ref{f2}) to annotate each
point of the given trace with connectivity information. The third
phase calculates the semantics (Figure~\ref{f4}) of the trace. The
fourth phase uses the connectivity information and the optimization
rules (Figure~\ref{f3}) to reduce the trace. The fifth phase
calculates the number of context switches in the resulted trace. The
last phase calculates the semantics of the resulted trace.
Calculating the number of context switches and semantics before and
after transformations makes \textit{BinTrcRed} transparent to the
programmers.

Four common multithreaded Java benchmarks were the subject of our
experiments. The first benchmark, CTSP, is a multithreaded solution
for traveling salesman problem using a concurrent bound and branch
algorithm. The second benchmark, CPhilo,  simulates the famous
dinning philosophers problem. The third benchmark, CWebDow, is a
multithreaded tool for downloading from servers and servers
reflection. The fourth benchmark, CMerge, is a multithreaded version
of the merge sort algorithm. The experiments were run on a Windows 7
system whose processor is Intel(R)-Core2(TM)-i5-CPU-(2.53GHz) and
whose RAM is 4GB.

\begin{table*}
  \centering
  \begin{tabular}{|l|l|l|l|l|l|l|l|l|l|}
  \hline
    &  LC & TC &  SR$_b$& CR& TR& SR$_a$& CS$_b$& CS$_a$     \\
   \hline
       CPhilo &81 & 6&  4.0 ms & 5.0 ms & 5.0 ms &  3.0 ms &  54 &  8  \\
    \hline
      CMerge &519 & 18&  25.0 ms & 29.0 ms & 32.0 ms &  26.0 ms &  541 &  93 \\
    \hline
    CTSP & 709 & 5&  68.0 ms & 78.0 ms & 97.0 ms &  54.0 ms &  9617 &  1143  \\
    \hline
      CWebDow &35175 & 3&  43.0 ms & 48.0 ms & 42.0 ms &  33.0 ms &  144 &  21  \\
    \hline
\end{tabular}
  \caption{Experimental results}\label{t1}
\end{table*}

The experimental results are shown in Table~\ref{t1}. For the sake
of accuracy, all information are averaged using results of 100 runs.
Parameters used to measure the performance are the following.
\begin{enumerate}
    \item LC: Numbers of lines in source programs.
    \item TC: Thread counts.
    \item SR$_b$: The semantics running-time before transformation.
    \item CR: Connectivity analysis running-time.
    \item TR: Trace-transformation running-time.
    \item SR$_a$: The semantics running-time after transformation.
    \item CS$_b$: The number of context switches before transformation.
    \item CS$_a$: The number of context switches after transformation.
\end{enumerate}

The following comments about results worth mentioning. It is noted
that the trace-transformation run-time (TR) is proportional to the
original number of context switches. The semantics run-time before
transformation is typically more than that after transformation.
This is justified with the reduction in the number of context
switches. The proposed algorithm managed to reduce number of context
switches by $85.3\%$ on average. This improves on the result of
\textit{SimTrace}~\cite{HuangZ11} whose average reduction percentage
is $83.8\%$. Compared to \textit{SimTrace}, the binary nature of our
proposed technique, \textit{BinTrcRed}, makes it more efficient for
larger traces. All in all, compared to the state of the art, these
results prove the value and usefulness  (regarding efficiency and
trace simplification) of the propped techniques. Two important
advantages of our proposed technique over related ones is that our
technique is supported with the operational semantics and a
correctness  proof for each trace transformation. The correctness
proofs have the form of inference rules derivations. This has many
applications; specially in the proof-carrying code area of research.

\section{Related Work}\label{s4}

Towards finding bug cases in error traces, many
algorithms~\cite{GroceCKS06,LeinoMS05} for checking software models
have been proposed. Most of these algorithms aim at building
counterexamples in case of finding a bug and aim also at reducing
error traces. Required changes in thread scheduling to get an error
that is concurrency-based was achieved by an extended version of
delta debugging~\cite{Artho11}. Assuming the existence of
rely-guarantee proofs for concerned properties, in~\cite{GargM11}
concurrent programs were verified. Although the proposed technique
in the current paper relies on producing reductions in single
traces, the techniques mentioned above rely on comparing related
traces. Clearly, focusing on reducing a single trace is more
practical and efficient but creates a more complicated scenario.

A static approach, \textit{SimTrace}, to  trace simplification is
proposed in~\cite{HuangZ11}. The idea behind \textit{SimTrace} is to
use dependence graphs to model events. Rather than introducing a
trace theorem for equivalence, in~\cite{HuangZ11} it is proved that
results of \textit{SimTrace} are sound. Hence in this approach
re-execution of program for the sake of validation is not required.
The use of a dependence relation~\cite{HuangZ11} is a common
practice in treating trace optimizations. Checking violations of
atomicity was achieved in~\cite{WangLGG10} via the introduction of
concept of guarded independence. To minimize the cardinality of the
causality relationship, the concept of sliced causality was
introduced in~\cite{ChenR07}. This was done by shopping the typical
dependencies among commands. Other research~\cite{SerbanutaCR12}
considered all possible valid executions that may result from a
trace. This was done using a model for maximal causality. The rule
of dependence relation is achieved in our proposed technique,
\textit{BinTrcRed}, by the connectivity analysis which is simpler
and more powerful than the mentioned techniques due to simplicity of
inference rules as they were explained earlier.


In~\cite{QadeerR05}, a theory of context-bounded analysis was
developed for concurrent programs. Up to the bound, this theory is
both sound and complete. Results concerning sequential pushdown
systems~\cite{BansalD13}, in particular their model checking, were
used to develop this theory. Many model checkers have been proposed
for concurrent programs~\cite{ChoDS13}. The problem with all these
checkers is that they use a representation of the stacks of threads.
Non-termination may occur due to such stacks. Other
techniques~\cite{GuptaPR11} for verifying concurrent programs that
are automated have also been developed. The idea in these techniques
is to use an automatically established model of the environment to
separately check each process. This checking model suffers from
being imprecise and stackless. Therefore such techniques are not
complete, but sound.

\bibliographystyle{plain}
\bibliography{Xbib}
%
%

\end{document}